In: Proceedings of the Eighth International Conference on Computer Supported Education (CSEDU 2016)

# Understanding Electric Current Using Agent-Based Models: Connecting the Micro-level with Flow Rate

Pratim Sengupta[1] and Uri Wilensky[2]

[1]Educational Studies in Learning Sciences, University of Calgary – Werklund School of Education, Calgary, Canada
[2]School of Education and Social Policy & Department of EECS, Northwestern University, Evanston, USA
pratim.sengupta@ucalgary.ca, uri@northwestern.edu

Keywords: Agent-based Models, Physics Education, Modelling, Computer Supported Learning, Mathematical Thinking

Abstract: Rate-based processes comprise an important set of scientific phenomena, as well as an important part of the K12 science curricula. Electric current is one such phenomenon, which is taught in various forms from 4th – 12th grades. Research shows that students at all levels find electricity difficult to understand, and the difficulties persist even after classroom instruction. In this paper, we present a design-based research study and argue that interacting with multi-agent-based computational models based on the microscopic theory of electrical conduction, can enable 5th grade and 7th students to develop a deep understanding of electric current as an emergent process of flow in terms of its microscopic level entities and their attributes, by bootstrapping their repertoire of intuitive knowledge. We present a particular design strategy – representing electric current as a fictive and transient process of charge accumulation, without falling in previously reported traps of the "source sink" mental models – and show how this strategy was effectively implemented in the computational model as well as in the learning activities performed by the students. We identify the mental models that students developed through their interactions with the model, and show that after their interactions, students were able to provide correct, multi-level explanations of the behavior of electric current in a resistive circuit.

## 1 INTRODUCTION

A large body of research has focused on the content and structure of the initial conceptual knowledge of physics novices in the domain of electricity, which shows that students at all levels (middle school through college) find basic electricity, i.e., electric current, resistance and the behavior of linear electrical circuits, hard to understand (for a review, see Reiner, Slotta, Chi & Resnick, 2000). Misconceptions that stem from these difficulties have been regarded by several researchers as resistant to change due to instruction (Haertel, 1982, 1987; Cohen et al., 1983; Reiner et al., 2000). Furthermore, several scholars have also suggested that the inability to link the behavior of individual electric charges to the overall behavior of electric current and resistance in circuits is a major reason for students' difficulties in this domain (Bagno & Eylon,1990; Frederiksen, Gutwill & White, 1999).

In this paper, we focus specifically on the conceptual development of a central concept in introductory electricity in middle school students: understanding electric current as a rate of flow of electric charges. Researchers have shown that even college students find rate-based processes such as electric current and molecular diffusion very challenging to understand (Miller et al., 2006; Chi, 2005). On the other hand, representing electric current in terms of microscopic-level entities (electrical charges) and their properties and behaviors using computer simulations has been shown to be an effective instructional approach at the high school level (Frederiksen, Gutwill & White, 1999). Our focus here concerns whether such a pedagogical approach can enable much younger students to learn develop a deep understanding of electric current as a rate of flow of charges. We report a design experiment (Edelson, 2002; Cobb, Confrey, diSessa, Lehrer & Schauble, 2004), in which we investigate the affordances and challenges in designing and implementing such a learning environment for middle school students (5th and 7th grades), and how we overcame the challenges.

The learning environment we used in this study is NIELS (NetLogo Investigations in Electromagnetism; Sengupta & Wilensky, 2009). NIELS is a suite of simulations we developed in the NetLogo multi-agent-based modeling platform and programming language (Wilensky, 1999). NIELS models depict electrical conduction in linear circuits from an emergent perspective (Wilensky & Resnick, 1999; Resnick, 1994). In an emergent perspective, aggregate-level behaviors (e.g., the collective movement of charges) emerge from simple behaviors and interactions between many individual-level agents or entities (e.g., electrical charges, ions, etc.). NIELS simulations are based on the microscopic theory of electrical conduction (Drude, 1925), which is explained in detail in Section 2.2. In NIELS models, electric current and resistance in a circuit are modeled as aggregate-level phenomena that arise due to simple, rule-based interactions with thousands of individual-level objects or agents (such as electrons, atoms and ions that constitute the circuit).

From the epistemic perspective, we will identify both the affordances and challenges associated with adopting such a pedagogical approach for learning about electricity. The specific learning goal we focus on concerns being able to understand electric current as a rate of flow of electric charges (we explain the physical mechanism later in Section 2.2). We present a NIELS simulation of linear electrical conduction from a microscopic perspective, and demonstrate how we iteratively designed and appropriated the underlying model and the simulation in order to address the challenges we identified for learners in 5th and 7th grades.

## 2 THEORETICAL BACKGROUND

### 2.1 The Micro-Macro Link in Electricity Education

Several scholars have studied the nature of initial (pre-instructional) knowledge of novices in electricity education. For example, research shows that most novices typically reason that current coming out of the circuit is less than that going in (for a review of misconceptions in electricity, see Reiner, Slotta, Chi & Resnick, 2000). That is, when current coming out of the battery meets the resistor, it slows down, and/or some of it is "lost" in overcoming the resistance. This type of reasoning has been termed as "current as an agent" model (White & Frederiksen, 1992), as well as "sequential reasoning" or the "current wearing out model" (Dupin & Joshua, 1987; Haertel, 1982). According to Chi and her colleagues (Chi, 2005; Slotta & Chi, 2006; Reiner et al., 2000), such misconceptions are generated due to novices' usage of "object-based" or "substance-based" knowledge, a coherent knowledge structure that includes ontological attributes of objects such as "being containable", "being pushable", "storable", "having volume" and "mass", "being colored', etc. They argue that experts think of electrical phenomena in terms of "process schemas" (Chi, Slotta & Leauw, 1994) or "emergent processes" (Chi, 2005; Slotta & Chi, 2006). Based on this argument, Chi and her colleagues have advocated that it is only through discarding the naïve ontology of electric current as a substance, and replacing it with the expert ontology of electric current as a process, that one can engender expertise in novices. They have argued for fostering such radical conceptual change through direct instruction focused on "ontology training", i.e., teaching the "process based" or "emergent ontology" (Chi, Slotta & Leauw, 1994; Slotta & Chi, 2006).

The central learning goal of the participants in this study reported in this paper involves them being able to understand electric current as an emergent phenomenon (a detailed explanation of the underlying physics is provided in Section 2.2). However, we believe that in order to understand such emergent processes, students' agent-based reasoning (i.e., students' intuitive knowledge about the individual-level entities or agents), as well as being able to conceive an emergent process in terms of discrete events, can play an important, productive epistemic role, and part of our agenda here is to highlight this role the process of students' conceptual development. Prior research in physics education provides support for our belief. For example, researchers have argued that the inability to link the agent-level behavior – i.e., behavior of individual charges – to the overall aggregate-level phenomena (i.e., electric current and resistance) that arise from the aggregation of the individual-level behaviors, is a major reason for students' difficulties in understanding electrodynamics (Bagno & Eylon, 1990; Frederiksen, Gutwill & White, 1999). This has been noted as the "missing micro-macro link" in electromagnetism education (Bagno & Eylon, 1990).

Researchers have found that while high schools students and college freshmen find electrostatic behavior easy to understand even prior to instruction, they find electrodynamics very challenging even after instruction. In order to bridge this divide, Eylon & Ganiel (1999) proposed an instructional module for high school students and college freshmen which reorganizes concepts in both electrostatics and electrodynamics based on underlying equation-based physics principles. Other researchers have showed that a combination of computer-based microscopic models of electrical conduction and algebra-based flow equations can indeed foster productive conceptual change in high school students regarding the behavior of electrical circuits (White, Federiksen, & Soephr, 1993; Frederiksen, White & Gutwill, 1999).

At the college level, Chabay & Sherwood (2004) have outlined a freshman-level college course that integrates theoretical microscopic models of electrical conduction along with traditional equation-based approaches for analyzing and understanding circuit behavior. Sengupta & Wilensky (2009) have proposed instructional modules based on multi-agent-based models that also seek to bridge the missing micro-macro link, and can be used successfully by freshmen students to develop conceptual understanding of electric current and voltage. Central to Sengupta & Wilensky (2009) and our current paper, is a construcvitist perspective (Smith, diSessa & Roschelle, 1994) grounded in a knowledge-in-pieces approach (diSessa, 1993), where in contrast to the work of Chi and her colleagues, students' fragmentary, object-based ideas can be successfully bootstrapped to develop deep understandings of electric current.

## 2.2 Potential Challenges in Understanding Microscopic Representations of Electric Current for Middle School Students

So far, we have argued for the affordances of adopting a microscopic perspective in electricity education. However, in the context of learning to conceptualize electric current as a "rate" while grounded in this perspective, middle school students (5th and 7th graders) may also face significant challenges. One source of these difficulties is their prior learning experience. For example, in the school where NIELS was implemented, "rates" are introduced to students in the latter half of the 7th grade academic year in their math classes, as is typical in many US schools. Therefore, at the time when NIELS was introduced to them, neither 5th nor 7th grade participants had studied rates before. On the other hand, high school students (12th graders) are already very familiar with "rates", as a part of their regular math and science curricula. Furthermore, 5th and 7th graders also have limited to no experience with algebraic equations. This suggests that understanding electric current as the "rate" of electron flow might be easy for 12th graders, but might prove to be challenging for 5th and 7th graders.

Another source of students' difficulty results from the possibility of conflicting predictions that might seem confusing to students. To understand this potential difficulty, let us undertake a closer examination of the microscopic theory conduction. At the heart of Drude's theory is the notion of free electrons, which are the electrons in the outermost shell of a metallic atom. When isolated metallic atoms condense to form a metal, these outermost electrons wander far away from the parent nucleus, and along with other free electrons, form a "sea" or a "gas" of free electrons. The remaining "core" electrons remain bound to the nucleus and form heavy immobile ions. In absence of an electric field, collisions with these ionic cores give rise to a random motion of the electrons. When an electric field is applied to this "gas" of free electrons, the electrons try to move against the background of heavy immobile ions towards the battery positive. It is the aggregate effect of these electron-ion collisions that give rise to electrical resistance, whereas electric current is the net flow of electrons resulting from the aggregate motion of individual free electrons (Ashcroft & Mermin, 1976; pp 24 – 49).

Let us now consider "n" free-electrons in a unit volume of a wire, each moving with a velocity "v". Then, in time t, each electron will advance by a distance v * t in the direction of its velocity. So, in time t, the number of electrons that will cross a unit area perpendicular to the direction of flow would be equal to n*v * t. Since each electron carries a charge e, the total charge crossing this unit area is equal to n * e * v * t. Now since electric current per unit area can defined as the number of electrons flowing through that area per unit time, therefore electric current can be expressed as:

$$I = \frac{n * e * v * \Delta t}{\Delta t} = n * e * v$$

(Source: Aschcroft & Mermin, 1965, pp 54)

The following multiplicative proportionality indicated by this equation is at the heart of the design of the NIELS models: the two cases of 800 electrons moving at 5 miles per hour, and 8 electrons moving at 500 miles per hour will result in the same amount (or value) of electric current. That is, according to this equation, electric current is represented as a rate (i.e., number of electrons flowing per unit time), as well as a quantity that can be conserved under the opposing influences of number and speed of electrons. Dealing with such conflicting predictions about number and speed is a central component of our designed learning activities.

Inhelder and Piaget (1985) and Siegler (1976, 1981) found that young children as well as adults have difficulties in reasoning about conflicting predictions in several contexts such as volume conservation, balance beams (and seesaw), etc. If a weight is placed on each side of the fulcrum in a balance beam, the beam will either tilt counterclockwise, or tilt clockwise, or not tilt at all. Canonically speaking, the effectiveness of a weight in causing the beam to tip is determined by the product of the weight (w) and its distance from the fulcrum (d), a construct called the torque associated with the weight. If the total torque associated with the weights on each side of the beam is the same, the beam will balance; otherwise, the beam will tip to the side with the greater torque. Piaget found that in reasoning about this task, initially children focus only on weight of the objects placed on either end of the beam; eventually by age 14, children develop the ability to coordinate weight and distance in the balance beam problem, and are able to identify their compensatory relationship (Inhelder & Piaget, 1985). In fact, other studies have shown that such reasoning is challenging even for adults (Lovell, 1961; Siegler, 1976; Hardiman, Pollatesk, & Well, 1986).

We therefore hypothesized that the effect of simultaneous and compensatory co-variation in number and speed of electrons may be difficult for younger students (5th and 7th graders) to understand. For example, while a higher number of electrons may lead to higher current, lower speed of electrons would lead to lower current. If a situation involves both these conditions occurring at the same time, then students might face challenges in making predictions. In the next section, we show specifically how we addressed this issue through redesigning the NIELS Current in a Wire model.

Our study has pragmatic significance. National science education standards in the US (NSES, 1996; NGSS, 2015) are premised on the argument that introducing microscopic level theories in 5th – 7th grade is premature. In contrast, we show that 5th and 7th graders do indeed have the intuitive repertoire to understand electrical conduction from a microscopic perspective, and can use bootstrap these understandings using agent-based models to develop a deep understanding of electric current as a continuous flow.

## 3. LOWERING THE THRESHOLD

### 3.1 The "Current in a Wire" Model

This model illustrates how a steady electric current and resistance emerge from simple interactions between the free-electrons and atoms (ionic cores) that constitute the electric circuit. It shows how the proportionality based relationships between current (I), resistance (R), and voltage (V) emerges due to the interactions between individual electrons and atoms in the wire. According to Drude's theory, in the presence of an externally applied electric field, each electron is accelerates till it suffers a collisions with an atom. As a result of this collision, the electron loses its velocity and again has to accelerate from a new initial velocity of zero, immediately after the collision.

The variables in this model are total number of free electrons (total electrons), voltage and number-of-atoms. Additionally, students can also "watch" an individual electron, as well as "hide" the electrons and atoms from their view without affecting the underlying rules of interaction between them, so that they can focus only on the trajectories of individual electrons. The graph displayed in the model plots the instantaneous current vs. time by calculating how many electrons are arriving at the battery-positive per unit time.

In this study, we also used an earlier iteration of the Current in a Wire model, in which the atoms are invisible to the learner (Figure 1). In this model, instead by controlling the number-of-atoms, they can control the probability of electrons experiencing collisions as they move towards the battery-positive.

### 3.2 The Redesigned Electron Sink Model & Rational for Re-design

While the previous model is aimed at focusing the learner's attention on the process of movement of free electrons inside the wire, the goal of this model is to frame the motion of electrons in terms of a process of "Accumulation" inside the battery positive in order to help 5th and 7th graders understand the notion of electric current as a "rate" in an intuitive fashion. Based on the literature in multiplicative reasoning, our hypothesis was that such a reframing would enable students as young as 5th graders to interpret electric current is terms of how fast the electron-sink fills up – a qualitative and comparatively more primitive form of "rate" based understanding, which in turn can be further developed onto a more formal understanding of rate through scaffolding in successive models, as we show later.

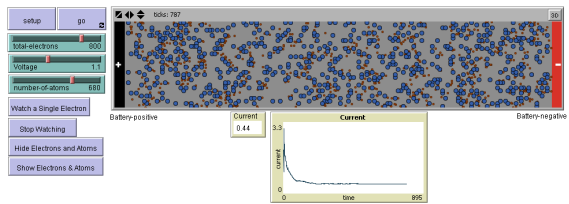

Figure 1: NIELS "Current In A Wire" Model

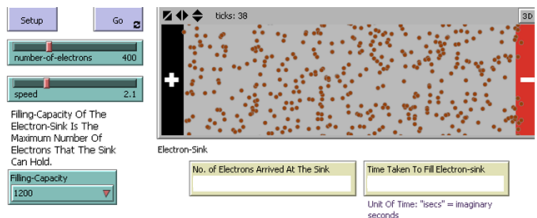

Figure 2: NIELS Electron Sink Model

In contrast to the Current in a Wire model, the Electron-sink model therefore has only two variables: number, and speed of electrons. The variables voltage and resistance that were present in the first model were condensed into a single variable, represented by the "speed" of electrons towards the battery positive. Another difference is that the atoms are also hidden from view. This was done so that students could focus on understanding the effects of the speed of electrons, as opposed focusing on the factors that control speed.

Our design rationale here is that once students are able to conceptualize electric current in terms of a process of accumulation – i.e., charges building up inside a contained environment - it would be easier for them to focus on the compensatory role of number and speed of the electrons being accumulated. Our rationale is based on research on children's development of multiplicative reasoning and part-whole relationships (compensation and covariation), where researchers have shown that novice learners can develop an understanding of rate by bootstrapping their intuitive knowledge about "building up" (Kaput & West, 1994). In terms of understanding compensation, of particular relevance is Piaget's study on class inclusion (Piaget, 1965). In his study, subjects were asked if a child who had four sweets to eat in the morning and four sweets to eat in the evening would have had the same number in total as the child who had been given one sweet in the morning and seven sweets in the evening. Concrete materials were available for representation. He concluded that children (7 years or older) could answer this question successfully by using their schema for compensation. As Irwin (1996) notes, in order to do so the children needed to be able to see a compensatory relationship between 4 + 4 and 7 + 1 in which 3 sweets were taken from one subset to the other. Presented symbolically, they needed to understand that P1 + P2 = (P1 + 3) + (P2 - 3). Some researchers have also suggested that children have a more sophisticated understanding of covariation and compensation that is protoquantitative, i.e., such reasoning can happen without exact quantification of entities, and that such early forms of understanding can potentially be bootstrapped successfully in classroom-based mathematics learning (Irwin, 1996; Resnick, 1992; Sophian & McCorgray, 1994).

The battery terminals are represented as "electron-source" and "electron-sink" in the user interface of the models, as well as in the activity sheets. This was done in order to prime students to tap into the semantic schemas of the terms "source" and "sink" and use them to interpret the functions of the battery-terminals. Finally, a third variable, "electron-sink-capacity", was introduced into the model. The function of this variable is to stop the model once a certain number of electrons reach the battery positive, and a "monitor" (see the right-hand side in Figure 2) displays the "time taken to fill the electron sink" (Ts). In terms of situational semantics, together with the "source-sink" metaphor, this creates an overall context of "containment", in which the electron-sink can be conceived of as a reservoir or a container that in which electrons are "building up" in number. Grounded in this broader context of containment and accumulation, some learning activities were designed specifically so that students could focus on the mutually compensatory role of the number and speed of electrons (see Activity D in Table 1).

Our framing of charge flow as accumulation is a fictive representation of the actual process of constant absorption of electrons by the battery-positive and regeneration of electrons in the battery-negative. The construct filling-capacity is also a fictive representation, designed to act as a scaffold to help students measure the rate of accumulation in terms of number and speed of electrons, the factors that affect the rate. The use of such hypothetical, transient models of steady-state electrical conduction has been long argued for by researchers as effective pedagogical tools (Chabay & Sherwood, 2004; Haertel, 1982; Ferediksen, White & Gutwill, 1999; etc.). Moreover, both in this model, as well as in subsequent models in the complete NIELS curriculum as we reported elsewhere (Sengupta & Wilensky, 2008, 2010, under revision), in subsequent models, students focus on a more accurate representation of continuous charge flow, where by tracking the motion of individual electrons throughout the circuit, they can actually see that once an electron reaches the battery positive, it gets absorbed and another electron is emanated from the battery-negative, in order to maintain steady-state in the circuit.

## 4. RESEARCH QUESTIONS

The specific research questions we seek to address in this study are:
  1. How do 5th and 7th grade students conceptualize electric current as a flow in terms of the microscopic level attributes depicted in the Electron-Sink model?
  2. After interacting with the Electron Sink model, were students able to provide multi-level explanations of electric current?

## 5. METHOD

### 5.1 Setting & Study Design

The research design is a mixed method study carried out as a design experiment, and includes both clinical interviews and quantitative analyses. The participants in this study consisted of two groups of students. One group of students consisted 5th and 7th graders (one section in each grade) who interacted with the NIELS Current In A Wire model (Figure 1), and is referred to in this paper as the Pilot group. Another group consisted of 5th and 7th grade students (one section in each grade) who interacted with the Electron sink model (Figure 2), and are referred to here as the Electron-sink Group. Each group represented diverse urban classrooms.

It is important to note that in the spirit of design based research, the comparison between groups presented in this study was not planned a priori as a controlled study. Rather, our goal is to highlight some interesting differences in terms of the learning experiences and learning outcomes among different groups of students, when each group of students used a modified version of the same model. The activities performed by students in the different groups are shown in Table 1.

During these implementations, students interacted with the NIELS models in randomly assigned groups of two or three, and the same group composition was maintained throughout the length of the implementation. Each NIELS model is accompanied by Activity Sheets that contain some relevant content knowledge (such as a multiple-analogy based introduction to free-electron theory, descriptions of "variables" and other functional features of the model's user interface) as well as instructions to guide students' interactions with the model. Each student was also required to log her or his observations and describe them in detail in these sheets, independently of her or his partner(s).

Table 1: Activities conducted by students

| Pilot Group | Electron Sink Group |
|---|---|
| A) Change only the value of number of electrons<br>a. Predict how and why electric current would change.<br>b. Observe (and compare with prediction) the effect of this alteration on electric current.<br>B) Change only the effective speed of electrons towards the battery-positive by controlling voltage, and<br>a. Predict (and explain), the effect of this alteration on electric current.<br>b. Observe (and compare with prediction) the effect of this alteration on electric current.<br>C) Change only the effective speed of electrons towards the battery-positive by controlling number-of-atoms, and<br>a. Predict (and explain) the effect of this alteration on electric current. | A) Change only the number of electrons<br>a. Predict, along with mechanistic explanations how the "filling time"(T) would be affected;<br>b. Observe (and compare with prediction) how T is affected;<br>B) Change only the speed of electrons and observe how T depends on it;<br>a. Predict, along with mechanistic explanations how T would be affected;<br>b. Observe (and compare with prediction) how T is affected;<br>C) Find two widely different sets of values of Number and Speed for which T is identical.<br>a. Why do you think the electron-sink filled up in the same time (T) in the two cases? Explain your answer in detail.<br>D) Given that electric current can be understood as "how fast the sink fills up", how would you measure electric current in the model? |
| b. Observe (and compare with prediction) the effect of this alteration on electric current. <br> D) Explain whether electric current would be equal, higher or lower, if twice as many electrons moved twice as slowly | E) Explain, if electric current would be equal, higher or lower, if twice as many electrons moved twice as slowly. |

The activity sheets also contained frequent prompts for reflection in which each student was often asked to provide detailed mechanistic reasoning of relevant phenomena. Each section consisted of 20 students. Note that the interventions reported in this paper lasted two class periods in each grade, and each period lasted 45 minutes. We conducted semi-clinical interviews with randomly selected four students in each class in 5th and 7th grades while they were interacting with the models. These students were identified from the class rolls prior to the beginning of the implementation. Once instruction began, we informed these students that they would be observed and interviewed periodically by one of the researchers present in the classroom. They were asked to raise their hand once they completed each activity listed in Table 1, and then the researcher would come and interview them about that activity. Our goal here was to minimize disruption in students' engagement with the activities, as a result of the interviews. In these interviews, which were videotaped, students were asked to provide mechanistic explanations of relevant phenomena, and we would often ask further questions to clarify and/or disambiguate parts of their responses.

It is also important to clarify the role of the teacher in our study, including classroom and group discussions led by her. During the first day of the study, the teacher, along with the lead researcher, led a classroom discussion in which students were provided with a handout with an annotated image of the relevant NIELS model. The handout identified the corresponding elements of the NIELS model and a diagram of a circuit by using arrows. The teacher had also setup a light bulb circuit on her table, and along with the researcher, led a classroom discussion in which she first asked all students to run the relevant NIELS model. She then asked several groups to identify, verbally, how the NIELS model represents the electrical circuit on her table. Once students started interacting with the NIELS model using the Activity Sheets, the role of the teacher and the second researcher was to respond to student queries about the user interface, and prompt students to explain their written responses in as much detail as possible, while the lead researcher was in charge of conducting interviews.

## 5.2 Analysis

In order to answer the first research question, we identify the Electron Sink group students' mental models and explanations about measurement of electric current, both during and after their interactions with the model. The individual case studies we present here are in the form of analysis of semi-clinical interviews of three students (Tara - 5th grade, Amber - 5th grade, and David - 7th grade). Our analysis of Tara's interview represents her mental model that emerged after she completed the Activity B in Table 1, and Amber's interview represents her mental model that emerged after she completed the second activity. David's interview was also conducted after he completed the second activity, but his mental model is different than Amber's mental model, and is representative of the diversity of mental models that we found among the interviews we conducted.

Our second research question seeks to answer the following question: After interacting with the Electron Sink model, were students able to provide multi-level explanations of electric current, when presented with a standard textbook representation of an electrical circuit devoid of any micro-level cues? In order to answer this question, we coded students' pre- and post-explanations of whether and why electric current should be equal throughout a light bulb circuit, along two dimensions: a) in terms of the commonly found misconceptions as noted by other researchers, and b) in terms of agent-aggregate complementarity. Along the first dimension, we found that students' incorrect explanations were typically in the form of "current decays as it moves through the wire". This type of explanation has been found in previous studies, and has been variously termed sequential reasoning (Fredette & Lockhead, 1980). This form of explanation was coded as Type A. The second type of response (Type B) was identified based on students' explicit mention of the closed nature of the circuit. These were correct explanations, in which students were able to identify that electric current remains constant throughout the circuit because of its closed nature.

Along the second dimension, as explained by Sengupta & Wilensky (2009), an aggregate-only perspective is an explanation that is devoid of any mention of micro-level agents and/or interactions.

Conversely, an agent-perspective would indicate an explanation that involves explicit mention of the individual-level agents and their interactions, without any explicit mention of an aggregate-perspective. An agent-aggregate complementary perspective would indicate an explanation that in addition to explicitly mentioning the agent-perspective, also describes how the aggregate-level phenomena emerges from the agent-perspective. In our case, explanations in which students explained the correct macro-level behavior (i.e., electric current is the same throughout the circuit) in terms of attributes and behaviors of micro-level agents (i.e., movement and/or circulation of electrons) were coded as agent-aggregate complementary.

In order to assess the reliability of our study, 20% of the pre- and post-tests and interview data were blind-coded independently by the first author of this paper and a second coder unaffiliated with the study. In both cases the coders agreed 95% of the time, resulting in a Cohen's Kappa of 0.9.

## 6. FINDINGS

### 6.1 Understanding Number-Speed Compensation

#### 6.1.1 Case 1: Identifying the relationship between movement and accumulation time.

This interview, as quoted in Excerpt 1, indicates that Tara, a 5th grade student, was able to identify that the filling time (T) is inversely proportional to the speed of the electrons. This interview was conducted immediately after Tara completed Activity B in Table 1. As shown in Excerpt 1, in lines 2 and 3, she explained that when electrons have higher speed, more of them reach the sink in less time. When the interviewer further asked her to clarify what she understood by T (lines 4 & 5), Tara clarified that she understood T to mean how fast the sink is filling up. When the interviewer wanted her to further elaborate the relationship between T and speed of electrons (lines 7 & 8), Tara explained the lowering of T in terms of simultaneous entry of more electrons inside the sink, and further compared the situation to "dropping more balls in the sink at one time" (lines 9 & 10). Our analysis therefore suggests that Tara interpreted the motion of electrons as depicted in the model both as movement within the wire, as well as accumulation (i.e., a building up process) inside the electron-sink, and that she was also able to explain a causal relationship between these two forms of movement - i.e., a higher speed inside the wire leads to a lower accumulation time in the sink.

Excerpt 1:
1. Int: So if the speed goes up why does T go down?
2. Tara: T goes down when the speed goes up because.. that means .. That means they are going through faster..
3. that means .. they are going to get there faster.. so T goes down..
4. Int: Cool.. umm.. so .. when you say T, do you mean T as in how much time electrons take to reach the
5. positive?
6. Tara: No.. it is how much time the sink takes to fill up..
7. Int: OK.. so I am still unclear why T would go down even if speed goes up.. I get it that electrons would
8. get there faster, but why would T go down?
9. Tara: Oh – that's because more is coming in.. at the same time.. its like you are dropping many more balls into
10. the sink at once.

#### 6.1.2 Case 2: Explaining number-speed compensation based on a proto-quantitative compensation schema.

In this interview, as transcribed in Excerpt 2, Amber (5th grader) explains how she got the value of T to be equal for two different sets of values of the number and speed of electrons. Amber's explanation here is based on the following: a) causal schemas that involve number and speed of electrons as individual causal agents, individually affecting the value of T; and b) a coordination of these causal schemas, i.e., simultaneous and mutually compensatory change in both the number of electrons and their speed, that can conserve the value of T. Amber's observations (as a part of Activities A and B in table 1), in the form of written responses, indicate that she was able to identify that number and speed of electrons, when varied individually, affected how fast the sink was filling up. In line 8, she explains that both these entities, when altered simultaneously in a compensatory manner, result in keeping the value of the filling-time (T) unchanged. To do so, she uses the following schema: in order to keep the sum (S) of the two numbers (A and B) constant, if one of the numbers is increased by a certain amount (x), then the other number must be decreased by the same amount (x). That is, $S = A + B = (A + x) + (B - x)$.

This indicates that Amber's explanation is based on a compensation schema (Piaget, 1965; Irwin, 1994). In this case, the "filling-time" is the result that remains unchanged when the values of both the number and speed are altered in a mutually compensatory way. This is reflected in line 10, when Amber explains that "you had to decrease their number, and umm.. the speed.. .. umm.. you had to increase, to make up for the time…"

Excerpt 2:
1. Int: so can you explain how you did the value of time to be equal in both cases?
2. Amber: what do you mean?
3. Int: I mean… the number (of electrons) here increased , and the speed decreased , right?
4. Amber: right..
5. Int: but you still get the same time, right?
6. Amber: right
7. Int: how is that happening?
8. Amber: well because you are taking off the 800 to get 500, and because you are taking off from one number, then you
9. have to add to another number.. like 4 + 3.. if you take 2 out of 4, then you have to put that two back on to 3… so it would be 2 + 5,
10. and that would be the same thing as 4 + 3… so that's what basically what we did…
11. Int: OK.. so that was a really nice explanation.. so could you explain that in terms of the wire and the electrons?
12. Amber: So.. umm.. for the electrons .. you had to decrease their number, and umm.. the speed.. .. umm.. you had to increase, to make up for the time…

### 6.1.3 Case 3: Explaining number-speed compensation using a "time-event" based mental model.

The following interview took place when David, a 7th grader, was performing the balancing time activity. The first author asked David to explain how they were planning to approach the task, and the conversation that ensued is transcribed in Excerpt 3. David's response in line 10 ("..number of electrons that were capable of being able to go into the positive charge at one time") indicates that he interpreted the situation as the one in which electrons were "going into" the battery positive. It indicates a process of simultaneous "entry" of multiple electrons ("going into", "at one time"), which we argue, plays an important role in David's explanation of how fast the electron-sink is filling up. David's written explanation quoted below provides a clearer picture of his mental model of the relationship between "filling time" and the process of simultaneous entry of electrons: "The filling time is lower when many electrons go in at the same time, so they fill up quickly. The time is the same if there are more electrons moving with less speed, and less electrons but moving faster".

Excerpt 3.
1. Int: … What's the idea?
2. David: So we are trying to get the speed.. trying to change the speed.. trying to variate .. kind of variate what the speed is…
3. 'coz we want to make it (pointing to "Ts" on the Activity Sheet) almost equal or close to 216 isecs, because in question 5,
4. that's what the speed is...
5. Int: So are you variating .. like increasing and decreasing the speed?
6. David: Yeah..
7. Sam (to David): well 2.6 is going to be too fast
8. Int: so can you tell me if the speed is going to be higher or lower?
9. Sam: We just got it.. it is 2.4..
10. David: Its (Ts) the exact same
11. Sam: Yeah.. we were able to get the exact same (Ts)
12. David (to the Interviewer): The speed was higher because the number of electrons that were capable of being able to go in to
13. the positive charge at one time was lower

Each "event", in David's explanation, represents a *unit* of time, during which a certain "number of electrons" enter the electron-sink. The notion of a event is evidenced in David's statement in line 10 in the above excerpt, where he explicitly mentions that a certain number of electrons "go in at one time". His explanation therefore indicates that the constituents of filling time are a series of these repeated, discrete time-events, each of which involves *simultaneous entry* of a multiple electrons. According to this model, if more electrons are capable of going in "at one time", the filling-time is lower, and vice-versa.

Furthermore, David's written explanation and interview excerpt also indicated that he was able to identify the mutually compensatory role of number and speed of electrons that resulted in conserving the value of filling-time. His interview response (see Lines 12-13 in Excerpt 3) indicated that a higher speed can compensate for a lower number of electrons, as it enables more electrons to enter "the positive charge (battery positive) at one time". This is also corroborated by his written response, where he identified that the filling time is the same if "if there are more electrons moving with less speed, and less electrons but moving faster".

## 6.2 Pre- and Post-Explanations of Electric Current in Light Bulb Circuit

Written responses in the pre-test indicate that 70% of 5th graders and 60% of 7th graders in Electron Sink group, and 75% of 5th graders and 70% of 7th graders in the Pilot Group indicated that electric current decays as it moves through the circuit. This type of incorrect responses was coded as Type A. Independent sample T-tests show that there is no statistically significant difference between the mean performances of the two groups, both in 5th grade ($t(19) = 1$, $p = 0.33$) and 7th grade ($t(19) = 1.4$, $p = 0.16$). However, in the Electron Sink group, in the post-test, only 10% of the responses in 5th grade and 15% of the responses in 7th grade were of this type. Paired sample T-tests revealed that the reduction in the percentage of these incorrect responses is statistically significant for both 5th grade ($t(19) = 4.49$, $p = 0.0001$) and 7th grade participants ($t(19) = 2.65$, $p = 0.02$). The second type of response (Type B) was identified based on students' explicit mention of the circuit topology, i.e., the closed nature of the circuit. Responses of this type correctly identified that electric current is the same everywhere in the closed circuit. 20% of the pre-test responses in 5th grade and 40% in 7th grade were of this type, while 80% of post-test responses in 5th grade and 75 % in 7th grade were of this type. Paired sample T-tests revealed that the gain in the percentage of these correct responses is statistically significant for both 5th grade ($t(19) = 3.26$, $p = 0.004$) and 7th grade participants ($t(19) = 2.51$, $p = 0.02$).

Another interesting statistic is the gain in the percentage of students' macro-micro complimentary explanations. Among Electron-Sink group participants in the 5th grade, 5% of students in the pre-test, and 60% of students in the post-test showed evidence of macro-micro complementarity in their written responses. Among Electron-Sink group participants in the 7th grade, 10% of students in the pre-test, and 60% of students in the post-test showed evidence of macro-micro complementarity in their written responses. In the pilot group, 10% of the students in 5th grade and 10% of the students in 7th grade showed evidence of macro-micro complementarity. Independent sample T-tests show that there is no statistically significant difference between the mean performances of Pilot and Electron Sink groups in the pre-test, both in 5th grade ($t(19) = 1$, $p = 0.33$) and 7th grade ($t(19) = 1.4$, $p = 0.16$). Paired sample T-tests also showed that the increase in percentage of macro-micro complementary responses among participants in the Electron-Sink group was highly significant, both in 5th grade ($t(19) = 4.49$, $p = 0.0003$) and in 7th grade ($t(19) = 3.59$, $p = 0.002$).

## 7. CONCLUSION & DISCUSSION

In this paper we problematized the educational design process itself. We reported a design experiment in which we highlighted design challenges in representing electric current as a process of flow in a manner that is aligned with the students' intuitive *sense of mechanism* (diSessa, 1993). We also provided a design strategy aligned with students' intuitive sense of mechanism in order to overcome the challenges: framing electrical conduction in terms of a fictive representation of accumulation of charges inside the battery-positive. The intuitive sense of mechanism, in this case, as we argued earlier, involves intuitive knowledge about accumulation, as well as appropriating the source-sink analogy at the microscopic level; and these in turn were bootstrapped by learners to develop an understanding of electric current as a continuous process of flow.

Earlier in this paper, however, we have also pointed out that many researchers have observed that novice students often use the "source-sink" analogy inappropriately in reasoning about the behavior of electric current in a circuit. As Reiner et al. (2000) points out, typically, such responses are also limited to the macroscopic-level descriptions of the relevant phenomenon. But, there is an important difference between the way in which source-sink models have been used traditionally, and in our approach. In our Electron-Sink model, one of the battery terminals acts as the source, whereas another acts as the sink. In contrast, the traditional use of the source-sink model involves thinking of the battery as the source, and the rest of the circuit (i.e. conductors, resistors, etc.) collectively as the sink. We also made explicit to the students the process of regeneration of charges, once electrons reached the battery negative, both visually (through the use of software scaffolds), as well as by teacher-led prompts.

The analysis presented in this paper suggests that the same source-sink model, when appropriated to describe the behavior of microscopic level objects (agents) such as electrons (instead of electric current, the macro-level phenomenon) can indeed act as a productive epistemic resource for students to develop a deep understanding of electric current as a process of flow in terms of its microscopic level entities and attributes. Furthermore, pre-post comparisons of the Electron Sink group's students' responses show that after interacting with the NIELS Electron Sink model, majority of them were able to develop explanations of electric current in a macro-micro complementary manner. This is consistent with Sengupta & Wilensky's (2009) finding that misconceptions about electric current and its behavior in linear circuits could be understood as evidences of ''slippage between levels''. That is, these misconceptions occur when students sometimes inappropriately assign the agent-level, substance-like attributes to the emergent phenomena, whereas these same knowledge elements, when activated at the agent-level description of the same phenomena, can lead to a deep understanding of aggregate-level processes.

Our analysis therefore shows that it is indeed possible for $5^{th}$ and $7^{th}$ graders to develop a deep, multi-level understanding of electric current as a process of flow in terms of the compensatory relationship between number and speed of electrons. Similar to Sengupta, Krinks & Clark (2015), our pedagogical design emphasizes cultivating learners' *sense of mechanism* (diSessa, 1993) rather than emphasizing a process of simple replacement of one idea with another. On a more general level, processes that involve aggregation comprise an important set of scientific phenomena across multiple domains – physics, biology, chemistry, materials science, etc. (Wilkerson & Wilensky, 2015). The design study we reported here shows how multi-agent based models can be designed so that they represent aggregation in a manner that is aligned with the students' intuitive sense of mechanism – by representing flow-rate in terms of discrete, dynamical events of accumulation, at the appropriate level of description (individual, as opposed to aggregate-level) of the phenomenon.